\documentclass[final]{svjour2}

\smartqed  
\usepackage{amsmath,amssymb}
\usepackage{geometry}
\usepackage{graphicx}
\usepackage{booktabs}  

\usepackage{color}  

\begin{document}

\title{Statistical crossover and nonextensive behavior of the neuronal short-term depression
}

\author{A.J. da Silva         \and
        S. Floquet         \and
        D.O.C. Santos
}

\institute{A. J. da Silva \at
          Centro de Forma\c{c}\~ao em Ci\^encias e Tecnologias Agroflorestais, Universidade Federal do Sul da Bahia, Itabuna, Bahia. CEP 45613-204, Brazil \\
              \email{adjesbr@ufsb.edu.br, adjesbr@gmail.com}           
           \and
           S. Floquet\at 
           Colegiado de Engenharia Civil, Universidade Federal do Vale do S\~ao Francisco, Juazeiro, Bahia. CEP 48.902-300, Brazil
           \and
           D.O.C.Santos\at
           Centro de Forma\c{c}\~ao em Ci\^encias e Tecnologias Agroflorestais, Universidade Federal do Sul da Bahia, Itabuna, Bahia. CEP 45613-204, Brazil   }

\date{Received: 2017 / Accepted: 10-2017}

\maketitle

\begin{abstract}
The theoretical basis of neuronal coding, associated with short term degradation in synaptic transmission, is a matter of debate in the literature. In fact, electrophysiological signals are commonly characterized as inversely proportional to stimulus intensity. Among theoretical descriptions of this phenomenon, models based on $1/f$-dependency are employed to investigate the biophysical properties of the short term synaptic depression. In this work we formulated a model based on a paradigmatic \textit{q}-differential equation to obtain a generalized formalism useful for investigation of nonextensivity in this specific type of synaptic plasticity. Our analysis reveals nonextensivity in data from electrophysiological recordings and also a statistical crossover in neurotransmission. In particular, statistical transitions providesadditional support to the hypothesis of heterogeneous release probability of neurotransmitters. On the other hand, the simple vesicle model agrees with data only at low frequency stimulations. Thus, the present work presents a method to demonstrate that short-term depression is not only governed by random mechanisms but also by a nonextensive behavior. Our findings also conciliate morphological and electrophysiological investigations into a coherent biophysical scenario.  
\end{abstract}

\section{Introduction}
\label{intro}
Neural communication is an intricate molecular process still not well understood. Information processing in the central nervous system (CNS) is mainly achieved by specialized structures called chemical synapses. Synaptic transmission is mediated by one or more neurotransmitter substances,  accomplished in the following steps \cite{wu,sudhof,ralf}: (1) action potential triggers opening of voltage gated calcium channels in the nerve ending; (2) opening of these channels allows influx of calcium ions into the neuron terminal; (3) on the active zone (AZ) of the cell membrane, calcium ions triggers vesicle fusion and neurotransmitter release into the synaptic cleft; (4) secreted neurotransmitters diffuse into the synaptic cleft, reaching receptors located in the postsynaptic neuron. Postsynaptic excitatory or inhibitory current ($I_{PSC}$) or potential ($V_{PSP}$) are prompted by neurotransmitters bound to the postsynaptic receptors. These electrical events are readily assessed by electrophysiological measurements. However, sustained presynaptic activity does not necessarily release the same amount of neurotransmitter into the synaptic cleft. Within the synaptic terminal, vesicles share a crowded environment, forming the readily releasable, recycling and reserve pools. These pools are successively recruited under sustained presynaptic stimulation, which initially promotes fusion of readily releasable or docked vesicles on the AZ. Higher frequencies promote a release probability increment of recycling and reserve pools, respectively. In other words, there is recruitment of vesicles from both pools toward synaptic fusion. 

The ability of neurons to change their vesicular dynamics, affecting synaptic strength, defines neuronal plasticity. For instance, a particular form of neuroplasticity, known as short term depression (STD), exhibited by different synapses in the brain, is characterized by $I_{PSC}$ or $V_{PSP}$ amplitude decrement. This promotes a degradation of synaptic transmission temporal fidelity, controlling the statistical properties of neurotransmission \cite{zucker}. To explain this mechanism, models for STD characterization based on $1/f$-dependency were developed to be tested over different experimental paradigms \cite{neher2}. Nevertheless, such models show limitations in accurately explaining experimental data. For example, studies show that neurotransmitter release does not behave like a haphazard process. Therefore, it is evident the need to develop more robust theoretical strategies. In this context, a possible nonextensivity in STD certainly contributes to clarify complex mechanisms involved in synaptic transmission \cite{neher,bennett,wernig}. Bernard Katz and colleagues introduced a statistical pillar for neurotransmission, quantifying the vesicular fusion as a Gaussian phenomenon \cite{bennett2}. Additional reports expanded this statistical description after considering other distribution functions \cite{robinson}. Relative to STD, inspite of its limitations, binomial statistics is the conceptual basis to characterize the degradation of plasticity. However, the exploration of statistical heterogeneity in neuroplasticity has not been contemplated using Tsallis Statistics. We hypothesize that the possible existence of statistical transitions and nonextensivity can overcome the restrictions of previous models by providing a more general scenario. 

Nonextensive Statistical Mechanics (NSM) describes systems in which the entropy is not proportional to the system size, a property frequently observed in complex systems that display long-range interactions or that are out of equilibrium \cite{tsallis88}. In this framework, Tsallis Statistics is successfully employed to investigate a variety of phenomena due to its ability to model power law phenomena \cite{tsallis2}. Although its widespread application, NSM remains scarcely applied in studies of the physiology of neurotransmission, despite the confirmation of nonextensivity associated to spontaneous release at the mammalian neuromuscular junction \cite{adr1,adr2}. However, our previous reports included neither brain synapses nor the possible role of electrical stimulation on nonextensivity. To overcome these limitations, using electrophysiological results collected at different synapses, we investigate whether there are nonextensivity and statistical transitions governing the neuronal communication involved in STD mechanism.

\section{Methods}
\subsection{Experimental data}
\label{Methods1}

Before introducing the theoretical analysis, we justify the use of the selected experimental data (see references for detailed experimental procedures). Electrophysiology, represented by a family of empirical tools, is largely employed to investigate neuronal activity, used in clinical examinations and in high-throughput screenings. Electrophysiological recordings are regularly applied for \textit{in vitro} studies, where patch-clamp and extracellular field potential techniques are employed to understand the substrate of neuronal plasticity. Among the preparations used in STD studies, one can highlight the auditory and limbic systems. Located in the auditory brainstem, the synapse formed by the calyx of Held and the main neuron medial nucleus of the mammalian trapezoid body (MNTB) is an important preparation to study STD due to their large cell size that facilitates empirical manipulations. Additionally, using the same morphological argument, the avian bulb of Held and the \textit{nucleus laminaris} make these models suitable for electrophysiological recordings, including STD assessment \cite{henrique,truss}. Moreover, the avian endbulb of Held and the \textit{nucleus laminaris} are relevant to address important questions in evolutionary neuroscience and comparative physiology of synaptic plasticity. Thus, these preparations expand our findings beyond those computed in mammalian species \cite{trussel}. The hippocampus, part of the limbic system, is a crucial brain area responsible for spatial memory, learning and navigation. A empirical advantage is the straightfoward process of tissue extraction, which allows cytoarchitecture preservation and accurate visualization of its different areas, being a highly used preparation for neuronal electrophysiology  \cite{nicoll,walmsley}. Adoption of the same formulation among different synapses is also justified, despite of functional and morphological particularities involved, due to common features in exocytosis machinery dynamics and similar electrophysiological response in STD curve. This strategy also allows to investigate how ubiquitous nonextensivity is in the nervous system.

Summarizing, the data used here were collected from intracellular $I_{PSC}$ and extracellular $V_{PSP}$ studies carried out in auditory synapses and in the dentate gyrus of hippocampal slices \cite {henrique2,cook,markram,kilbride}. Importantly, in spite of the methodological differences between extracellular and intracellular measurements, there is evidence that intracellular electrophysiological properties can be predicted by extracellular recordings \cite{buz}. This argument supports applications of nonextensive analysis of intra or extracellular electrophysiological recordings.

\subsection{Theoretical modeling}\label{Methods2}
\subsubsection{Crossover Statistics}

Boltzmann-Gibbs statistics (BG) states that the entropy additivity law, only valid for extensive systems, is governed by $S = -k \int P(x) \ln P(x)$. In this case, the exponential probability density $P(x) \propto \exp(-x)$, represents the entropy distribution of noninteracting systems. 
Because it considers long-range correlations, NSM brings a generalization for the classical description, since in their foundations a nonextensive or nonadditive entropy rule is assumed for $S_{q}$, written as:  

\begin{equation}\label{qE1}
S_{q}(A+B) = S_{q}(A)+S_{q}(B)+(1-q)S_{q}(A)S_{q}(B)
\end{equation}
In this case $A$ and $B$ are two independent systems with:
 
\begin{equation}\label{qE2}
P(x,x')_{A+B} = P(x)_A P(x')_B 
\end{equation}

In this sense, $P(x)$ represents the probability density distribution of the macroscopic variable $x$. Therefore, the so-called entropic index $q$ expresses the magnitude of a nonextensivity operating in the system. The maximization of the $q$-entropy leads to:

\begin{equation}\label{qE3}
S_{q} = \frac{k}{(q-1)}\left( 1-\int [P(x)]^q dx \right), \quad q \in \mathbb{R} 
\end{equation}
Its optimization produces a $q$-exponential distribution:

\begin{equation}\label{qE4}
P_q(x) \propto e_q^x\equiv\left[1+\left(1-q\right)x\right]^{1/(1-q)} 
\end{equation}
if $1 + \left(1-q\right) x \geq 0$ and $e_q^x = 0$ otherwise. 
In the limit $q \rightarrow 1$ the usual BG entropy is recovered, $S_1\equiv S_{BG}$, 
and Eq. (\ref{qE4}) converges to the usual exponential distribution, $P(x) \propto e^x$. Eq. (\ref{qE4}) can also be obtained from:
\begin{equation}\label{qE4a}
   \frac{dP}{dx} = -\lambda_{q} P^{q},  \lambda_{q} \geq 0; q\geq 1
\end{equation}
where $P = P_q(x)$ has a solution $P_q(x)=\left[1+\left(1-q\right)\lambda_{q}x\right]^{1/(1-q)}$. This power law, exactly the same $q$-exponential function showed in Eq. (\ref{qE4}), was applied to biological systems, discriminating supperdiffusive patterns in dissociated cells from \textit{Hydra} and giving a novel description of internucleotide interval distribution \cite{hidra,bunde}. 

We now discuss the conditions under which varying a macroscopic variable leads to the changing of a statistical regime governing the form of a probability density distribution. First, it is necessary to generalize Eq. (\ref{qE4a}) to one that unifies BG statistics and nonextensive statistics. This is accomplished by the paradigmatic equation \cite{tsallisbook}:
\begin{equation}\label{qE4b}
 \frac{dP}{dx} = -\mu_{r}P^r-{(\lambda_{q}-\mu_{r})}P^q,  \qquad   r \leq q, \quad  q \geq 1
\end{equation}
For $r = q = 1,\forall \lambda_{q}$ we recover BG statistics, that is $\displaystyle \frac{dP}{dx} = -\lambda_{q}P$, $q = 1$. For $\mu_{r} = 0, \forall r$ or if $r = q, \forall \mu_{r}$ we recover usual nonextensive statistics, given by Eq. (\ref{qE4a}). In this equation, $r$ measures the degree of both nonextensivity ($r > 1$) and extensivity ($r = 1$) in the same sense as the $q$-index. Solutions of Eq. (\ref{qE4b}) permit to observe a crossover from different statistical regimes from low $x$ values, dominated by a $q$-exponential behavior, to high $x$ values described by $r$-exponentials \cite{tsallis2,protein}. Eq. (\ref{qE4b}) was successfully employed to detect nonextensivity and to determine the statistical crossover in studies of the flux of cosmic rays \cite{cosmico} and protein folding \cite{protein}. Interestingly, under assumptions, this equation can also be seen as a generalization of Planck statistics \cite{beck}.  

Parameters $\lambda_{q}$ and $\mu_{r}$ determine the values of $x$ marking the change of a statistical regime. We observe this with a particular solution for $r = 1$ and $ q > 1$, which presents transition from nonextensivity to extensivity \cite{tsallis2,protein}:
\begin{equation}\label{qE4c}
 P(x)  =  \left( \frac{1}{1-\dfrac{\lambda_{q}}{\mu_{r}} + \dfrac{\lambda_{q}}{\mu_{r}}\exp \left[ (q-1) \mu_{r} x \right]}\right)^{1/(q-1)}  ,  \quad x\geq 0
\end{equation}
From this expression, the values of $x$ marking a given crossover are stated by the following expressions, for $r \neq q$, $\mu_{r=1} \ll \lambda_{q}$. In this case, we have a first crossover in $x^{*}_{q}$ and a second crossover in $x^{*}_{r=1}$:
\begin{equation}\label{cross1_r1}
 x^{*}_{q} = \frac{1}{\lambda_{q}(q-1)}
\end{equation}

\begin{equation}\label{cross2_r1}
 x^{*}_{r=1} = \frac{1}{\mu_{r=1}(q-1)}
\end{equation}

For $1 < r < q$, expressions for $x$ in a crossover can be found in \cite{protein,cosmico}. 
 
\subsubsection{Crossover Statistics in Short Term Depression}

The amplitude response of the electric signal involved in STD is characterized as inversely proportional to stimulus frequency. As previously discussed, many theoretical descriptions of this phenomenon were based on $1/f$-dependency, employed to investigate the biophysical properties of STD. Injection of a repetitive stimulation in the pre-synaptic terminal decreases $I_{PSC}$ or $V_{PSP}$ responses yielding STD. Let us consider a variable $R$ representing $I_{PSC}$ or $V_{PSP}$ responses. A formulation based on the $1/f$-behavior, used by other authors to investigate STD data \cite{neher2}, is denominated the simple vesicle depletion model given by the following expression:

\begin{equation}\label{qE5}
 R = \frac{1}{1 + f p \tau}
\end{equation}
where $\tau$ is a relaxation time constant toward a steady state, $f$ is the stimulation frequency and $p$ is the release probability. It is important to mention that this model neglects vesicle interactions into the synaptic terminal, being consistent with a binomial statistical description. We can further generalize Eq. (\ref{qE5}) giving it a power law format, introducing an exponent $n$:

\begin{equation}\label{qE6}
 R = \left( \frac{1}{1 + p \tau f}\right)^{n} 
\end{equation}

If we make $\displaystyle n \ = \ \frac{1}{q-1}$ and $ \displaystyle \tau p \ = \ \lambda(q-1)$, where $q$ is the nonextensive index, a nonextensive simple vesicle depletion model is obtained: 
\begin{equation}  \label{qE7}
  R =   \left( \frac{1}{1 + \lambda (q-1) f} \right)^{\frac{1}{q-1}} 
\end{equation}
Here, we also make the hypothesis that a $q$-exponential function governs $I_{PSC}$ or $V_{PSP}$ responses. As shown above, Eq. (\ref{qE7}) is a solution of the following: 

\begin{equation}\label{qE8}
   \frac{dR}{df}  =  -\lambda R^{q} 
\end{equation}

A further generalization of Eq. (\ref{qE8}) is possible if we introduce another nonextensive index, $r$ and assume that nonextensivity and statistical crossover occur in STD. This equation is similar to Eq. (\ref{qE4b}):

\begin{equation}\label{qE9}
 \frac{dR}{df} = -\mu R^r-{(\lambda-\mu)}R^q
\end{equation}
With $r \leq q$ and $q \geq 1$. Eq. (\ref{qE9}) admits an analytical solution as a family of hypergeometric functions, which give an approximated solution as they have to be truncated \cite{protein}. For this reason, we choose to numerically integrate Eq. (\ref{qE9}) to investigate nonextensivity in STD. This equation also predicts that increasing frequency causes the changing of a statistical regime governing STD. For example, crossover frequencies, for $r \neq q$ and $r = 1$ are given by:
\begin{equation}\label{cross3_r1}
 f^{*}_{q} = \frac{1}{\lambda (q-1)}
\end{equation}

\begin{equation}\label{cross4_r1}
 f^{*}_{r=1} = \frac{1}{\mu (q-1)}
\end{equation}

\begin{figure}[!htb] 
\centering
\hspace{0.1cm} \includegraphics[width=6.0cm]{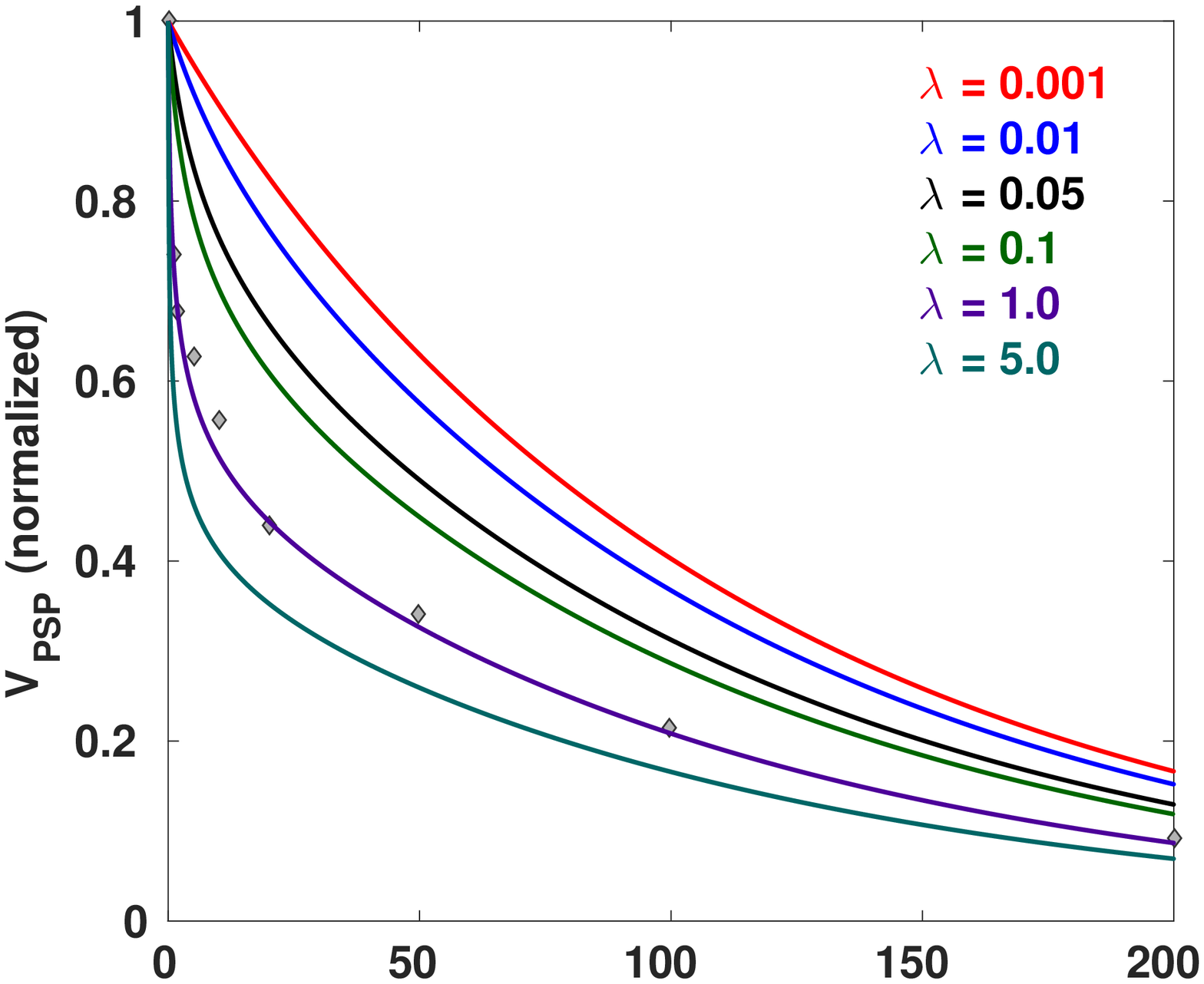}
\includegraphics[width=6.0cm]{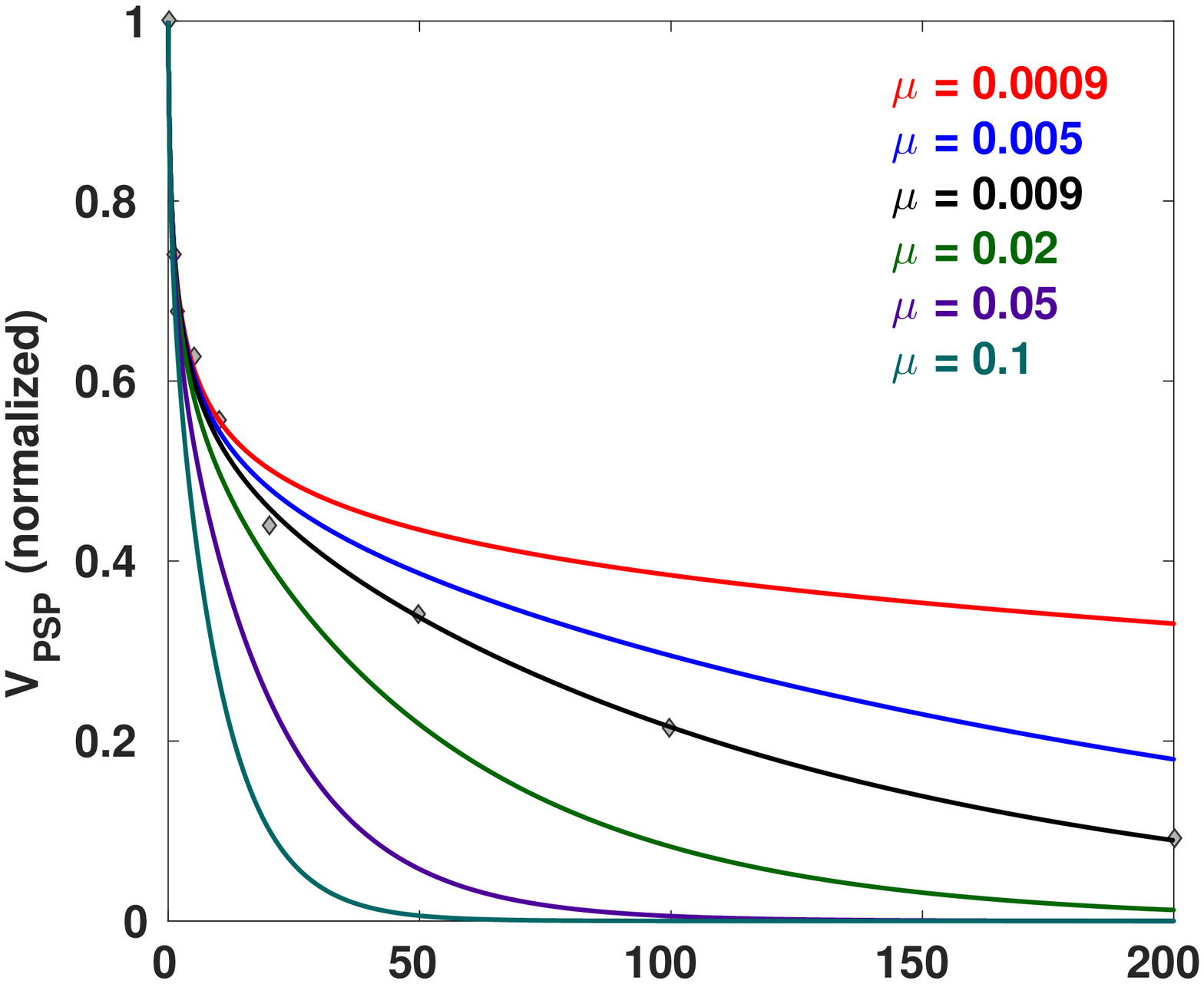}

\vspace{0.3cm}

\includegraphics[width=5.8cm]{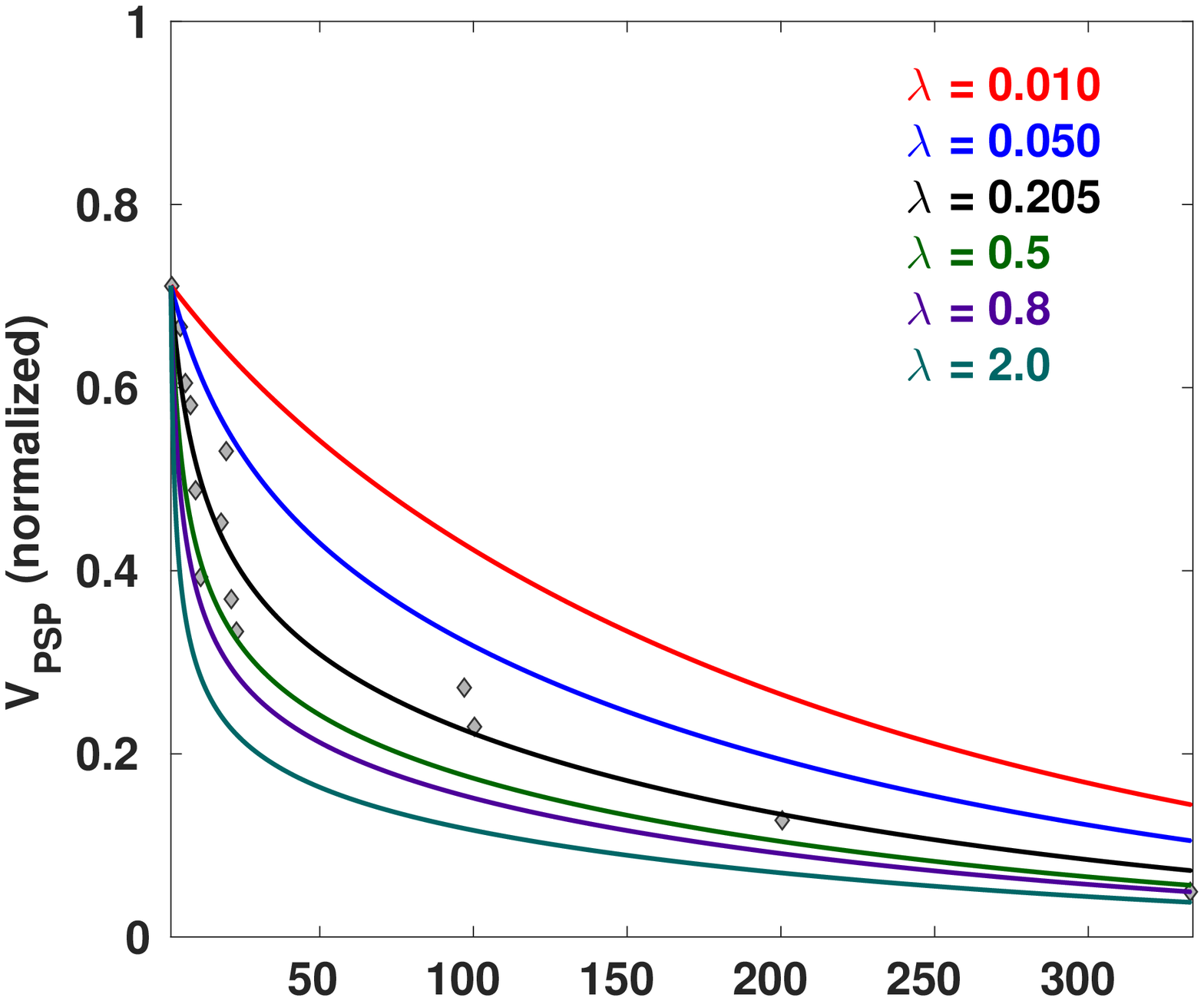}
\hspace{0.1cm}
\includegraphics[width=5.8cm]{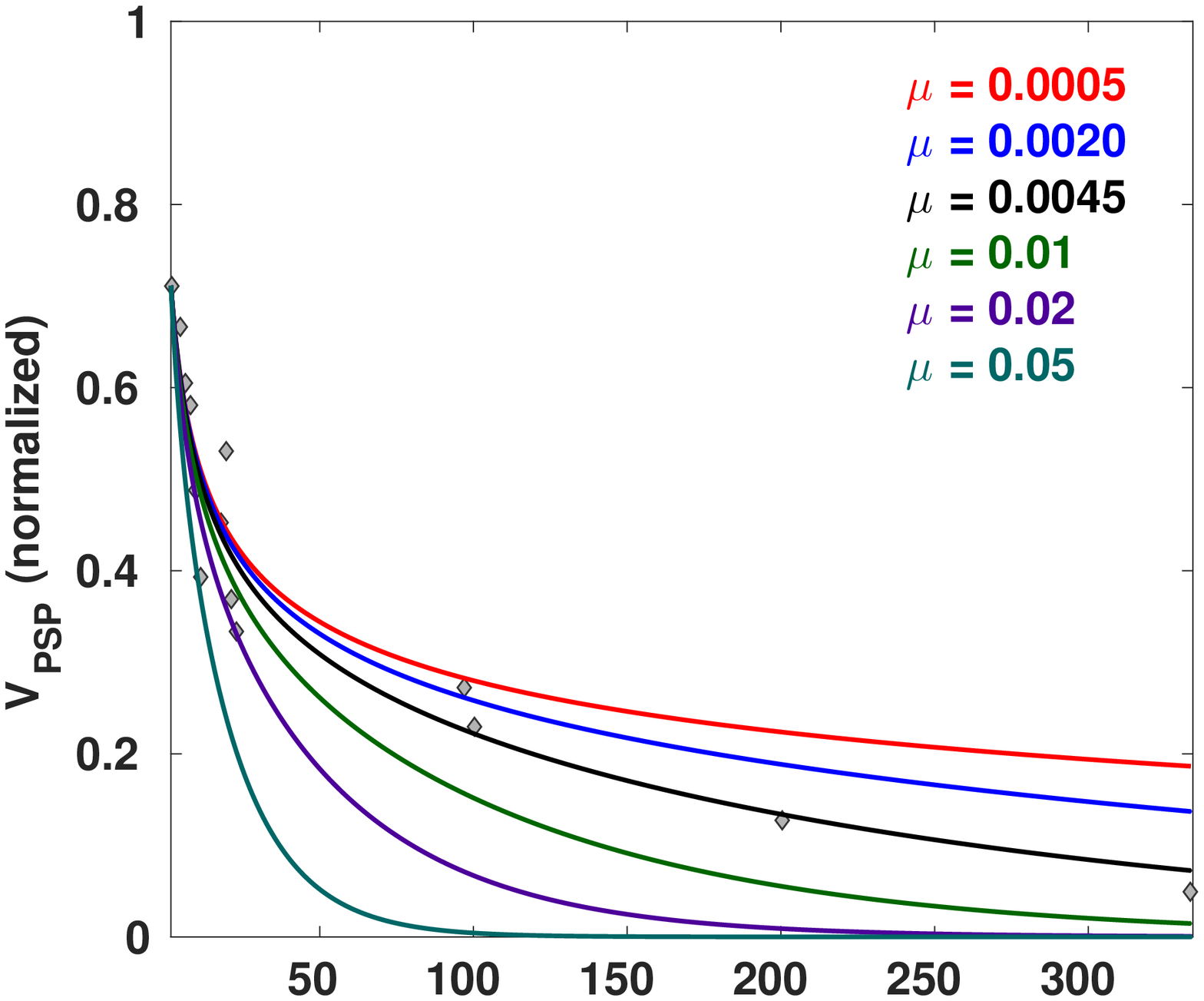}

\vspace{0.3cm}

\hspace{0.1cm}\includegraphics[width=6.0cm]{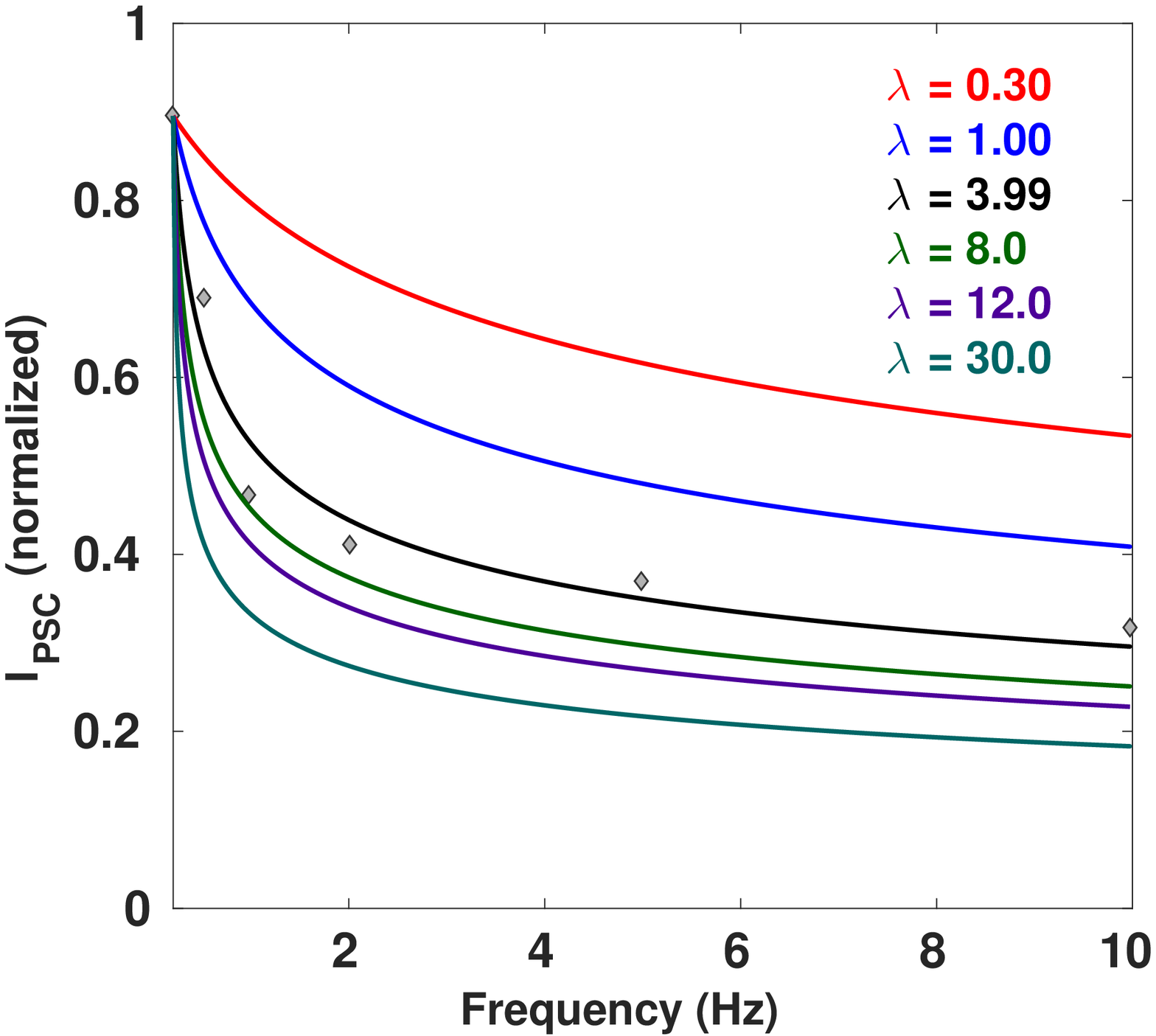}
\includegraphics[width=6.1cm]{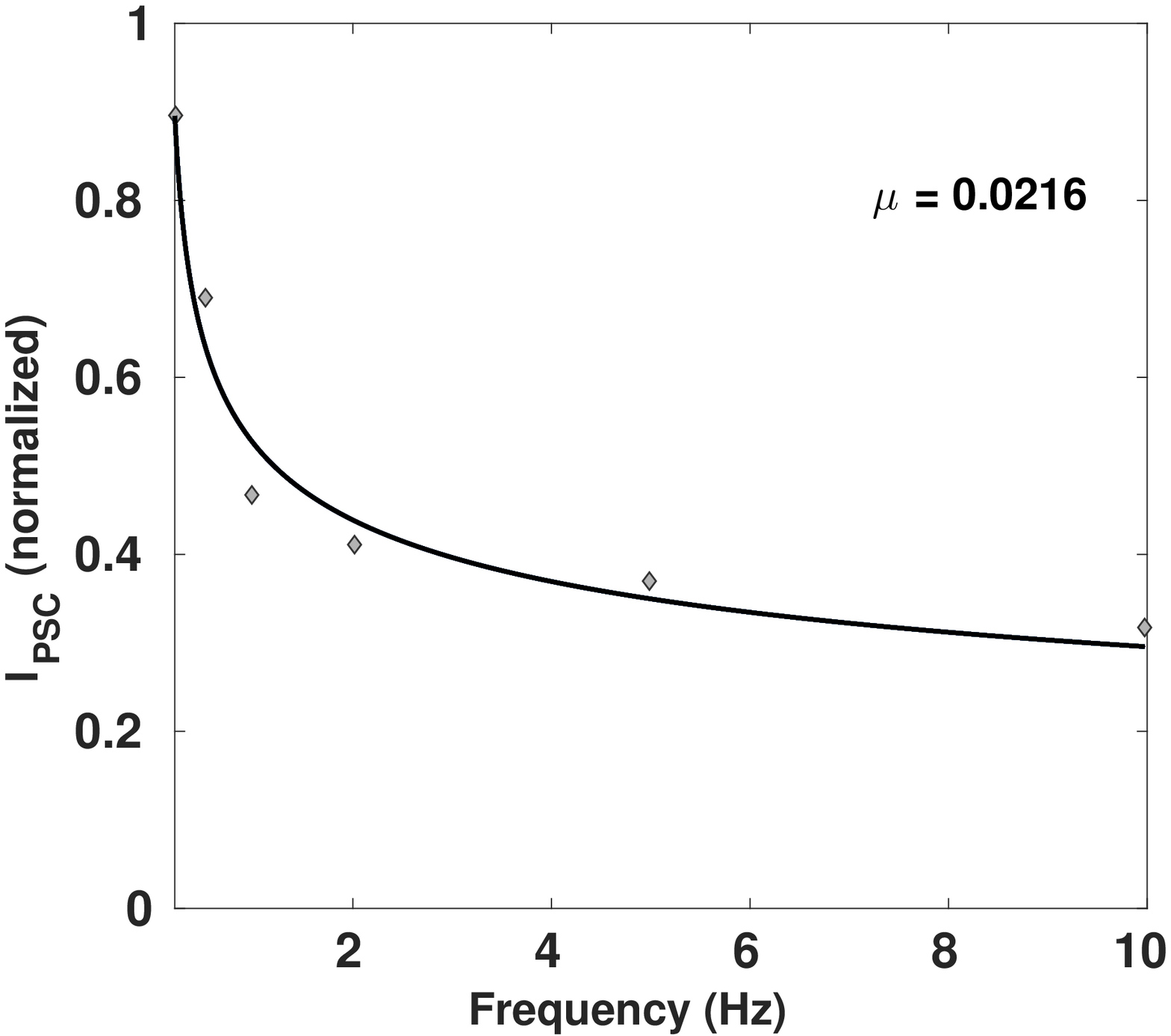}

\caption{Left: adjustments, with Eq. (\ref{qE9}), for several values for $\lambda$, keeping the other parameters fixed. Right: same as left, but for several values of $\mu$. Data points correspond to data from: \cite{kilbride} (top), \cite{cook} (middle) and \cite{neher2} (bottom). According to these simulations, $\lambda$ drives the rate of decay in the early phase (lower frequencies), while $\mu$ controls the degree of depression relative to the maximum $I_{PSC}$ or $V_{PSP}$.} 
\label{figure0a}
\end{figure}

To uncover how parameters $\lambda$ and $\mu$ change the shape of $R$ we performed simulations varying those parameters in Eq. (\ref{qE9}). The results are presented in Fig. \ref{figure0a}. Our simulations suggest that $\lambda$ adjusts the curve concavity or drives the rate of decay in the early phase (lower frequencies). Whereas $\mu$ controls the degree of depression relative to the maximum $I_{PSC}$ or $V_{PSP}$. This also suggest that $\mu$ and $p$ could be related. 

\begin{figure}[!htb]
\centering
\includegraphics[width=7.0cm]{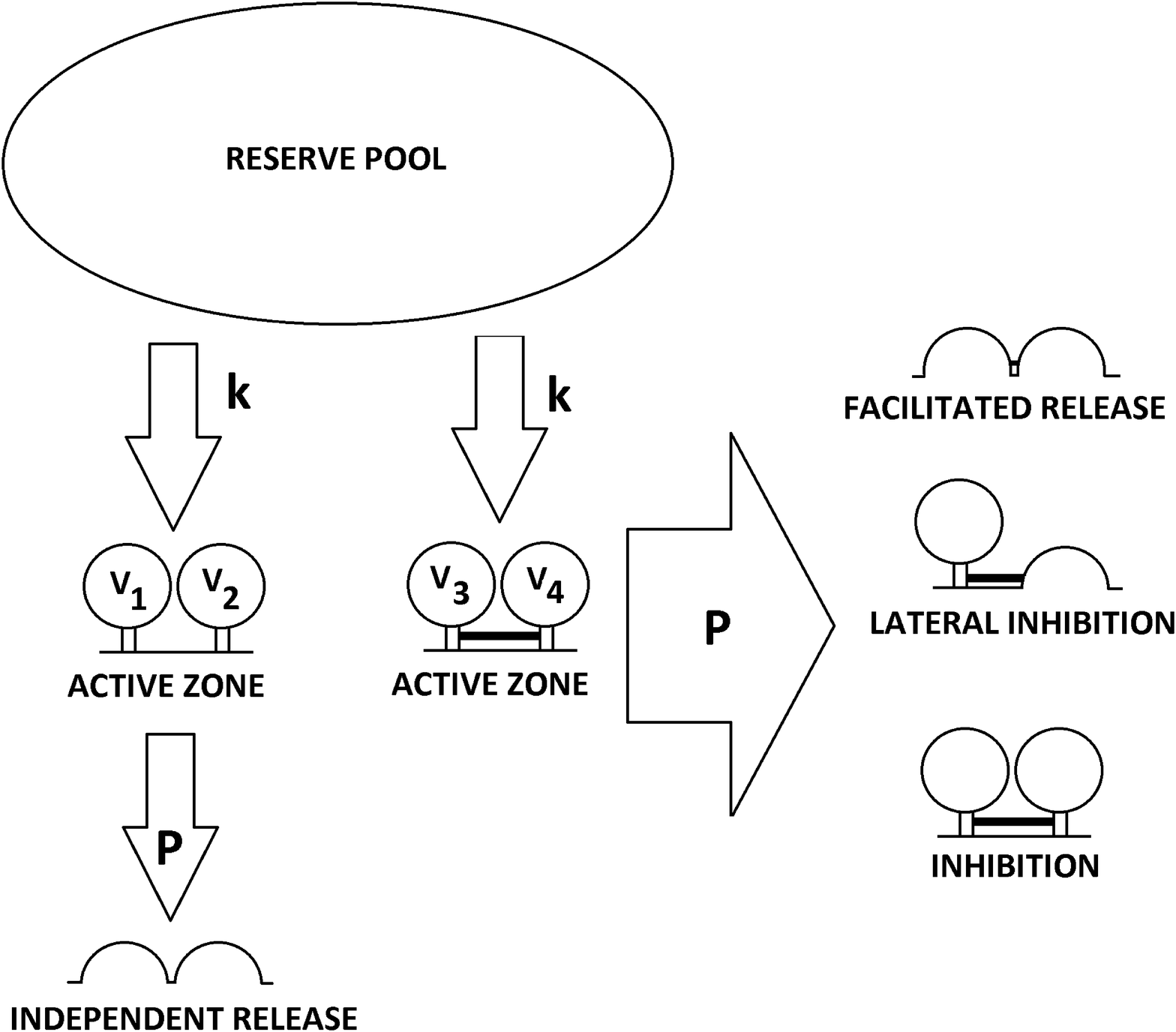}
\caption{Illustration of a STD scenario presenting nonextensive and extensive statistics. Vesicles are recruited from a reserve pool with a constant rate recruitment ($k$). On the AZ, interaction absence (V$_{1}$ and V$_{2}$) or presence (V$_{3}$ and V$_{4}$) are possible. The latter may emerge as a consequence of protein sharing from the SNARE (Soluble NSF Attachment protein REceptor) complex (represented by a black stripe), yielding three possibilities: facilitated release, lateral or partial inhibition or total inhibition. This is the substrate of nonextensive mechanisms. On the other hand, an independent release is connected to an extensive process.}
\label{figure0b}
\end{figure}

Given the information in Fig. \ref{figure0a}, it is necessary to write $\lambda$ and $\mu$ as functions of parameters related to STD. To achieve this and influenced by previous models, we start listing the main premises of our model: (1) the synaptic terminal is constantly supplied from an unlimited vesicle reservoir, where vesicles are recruited independently from each other, with rate $\kappa$; (2) vesicles, placed into the readily releasable pool at an active zone, are allowed to interact with each other by the proteins of the exocytotic machinery; (3) vesicles possess a mean residence time or a relaxation time given by $\tau$. (4) we do not take into account a refilling rate, that is, we are neglecting endocytosis contribution; (5) the release probability $p$ and quantal size $Q$ of vesicles are parameters independent of each other. Quantal size is the postsynaptic response originated from the release of a single vesicle; (6) there is a relationship between the $p$ and the $q$-index; (7) we do not consider calcium contribution. 

Finally, we suggest a non-extensive physiological mechanism depicted in Fig. \ref{figure0b}. According to this illustration, the exocytotic machinery comprises two types of mechanisms. Beyond the random mechanism, coherent with an extensive scenario, three situations emerge, as consequence of non-extensivity: lateral inhibition, facilitated release and total inhibition. 

We propose that $\lambda$ and $\mu$ are functions of the new parameters $\tau$, relaxation time; $\kappa$, recruitment rate of vesicles; $p$, release probability; $Q$, quantal size: 

\begin{equation}\label{qE12}
 \mu = pQ/ \kappa
\end{equation}

\begin{equation}\label{qE13}
\lambda - \mu = p Q \tau
\end{equation}
This implies $\mu > 0$ as this parameter is a product of positive quantities. Given these assumptions, we propose an equation for $R$, incorporating crossover statistics, which resembles one in the work by Niven \cite{niven}. The result is the $q$-differencial equation:

\begin{equation}\label{qE11}
 \frac{dR}{df} =- \frac{p Q}{\kappa} R^r- pQ\tau R^q
\end{equation} 
In this case, $r \leq q$ and $q \geq 1$.  Parameters $r$ and $q$ are crossover exponents as presented in the former section. It is important to stress that Eq. (\ref{qE9}) and Eq. (\ref{qE11}) are equivalent.

\subsection{Data analysis and optimization}

We used WebPlotDigitizer \cite{Rohatgi} to extract data from articles. Parameters from Eq. (\ref{qE9}) were estimated using genetic algorithms (GA), a class of optimization or parameter search algorithms incorporating biological evolution mechanisms \cite{Mitchell1996}. We used GA to find a vector in parameter space that minimizes the root mean squared difference between experimental data and simulated points from Eq. (\ref{qE9}). Computer simulations were performed in R-cran and MATLAB. Numerical calculations were run independently, in the sense that the researchers interacted with each other only after the parameters were obtained. Even with this strategy, our results were similar by themselves, which showed the robustness and reliability of our data analysis.

\section{Results}

\begin{figure}[!htb]
\centering
\includegraphics[width=7.8cm]{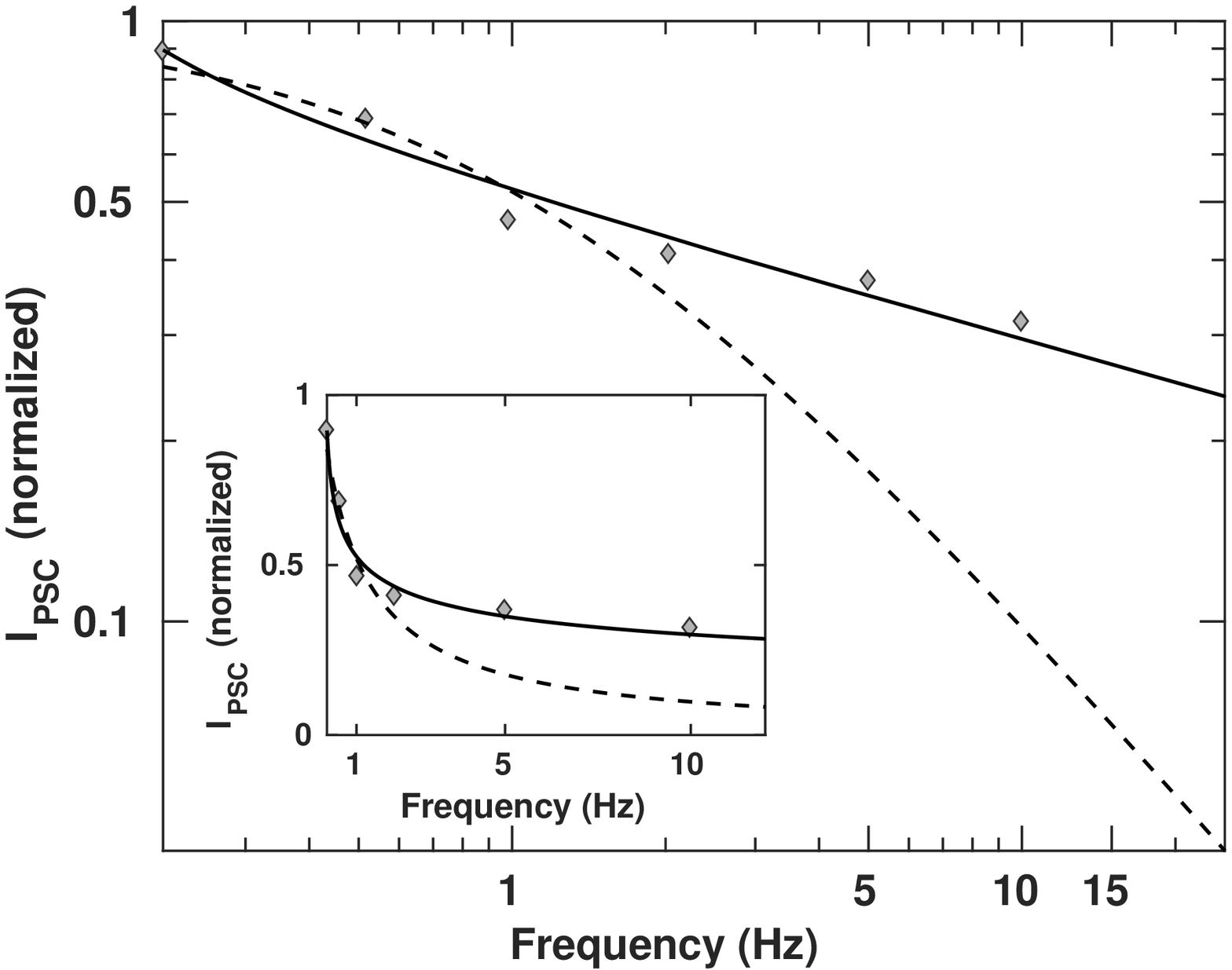}
\vspace{5mm}
\hspace{0.1cm} \includegraphics[width=8.1cm]{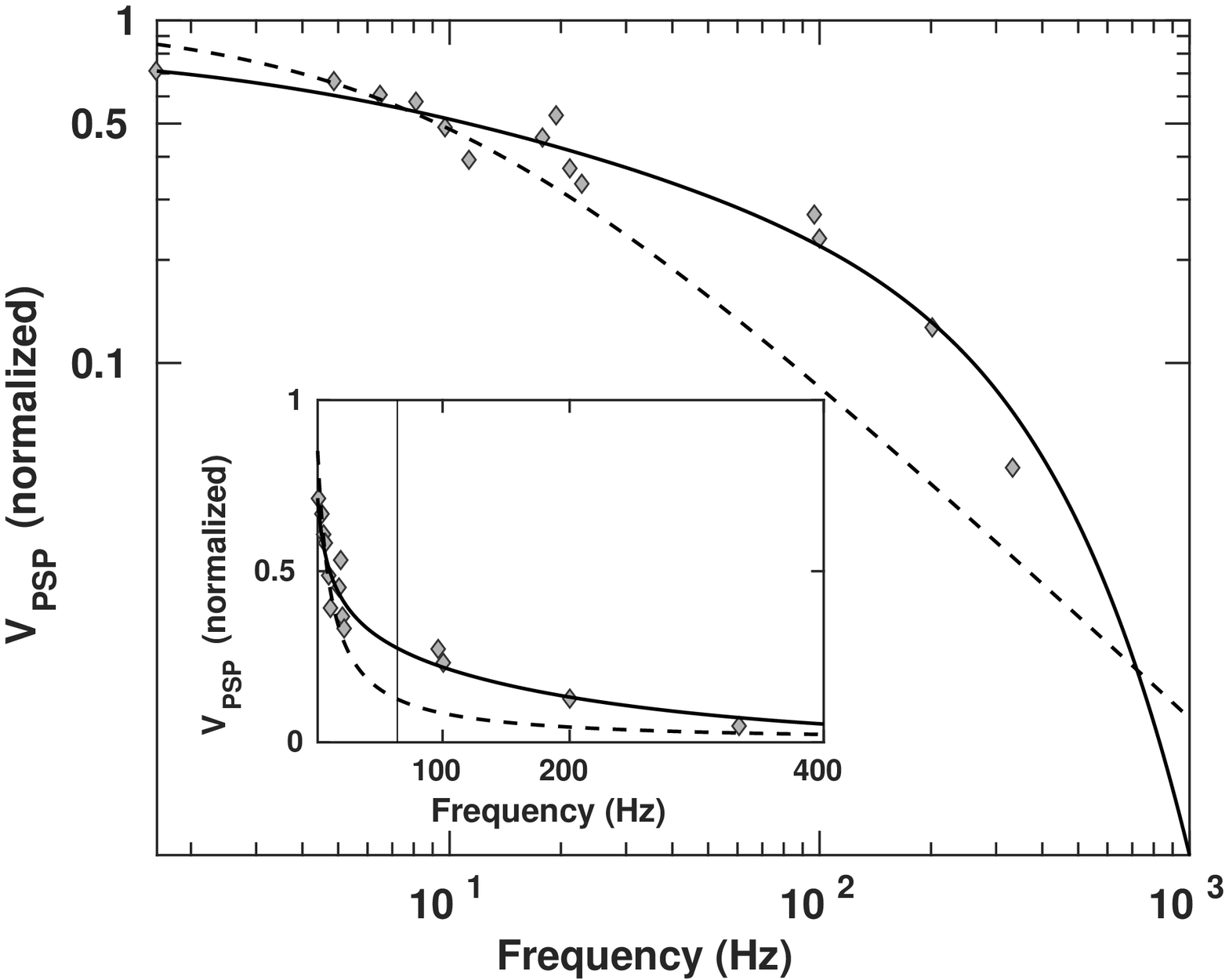}
\caption{Auditory system data recorded with patch clamp technique (log-log and linear-linear scales) and respective adjustments for both models. Full lines represent fitting with Eq. (\ref{qE9}), while dashed lines correspond to adjustments with Eq. (\ref{qE5}). Top: Data points adapted from excitatory $I_{PSC}$ recordings measured by von Gersdorff \textit{et al}.(\cite{henrique2}, Fig. 2A), also analysed in Weis \textit{et al}. (\cite{neher2}, Fig. 2B) and Trommershauser \textit{et al}. (\cite{trommershauser}, Fig. 4A). Bottom: Fits using excitatory $V_{PSP}$ data (\cite{cook}, Fig. 1D. The insets show the linear-linear representation of data. In the bottom inset, the vertical line marks the crossover frequency at 75.165 Hz. The first crossover at 1.467 Hz is not visible.}
\label{figure1} 
\end{figure}

We first applied Eq. (\ref{qE9}) to determine either \textit{q} and \textit{r} indexes and parameters $\lambda$ and $\mu$. We calculated the release probability, $p$ and $\kappa$ using Eq. (\ref{qE12}) and Eq. (\ref{qE13}) for fixed values of $Q$ and $\tau$ from the literature. Crossover frequencies were calculated using Eq. (\ref{cross3_r1}) and Eq. (\ref{cross4_r1}). In experimental protocols of STD, the calyx of Held synapse is characterized by a rapid $I_{PSC}$ decay followed by a pronounced steady-state region at higher frequencies. The adjustment of data from the calyx of Held (Fig. \ref{figure1}, Top from \cite{henrique2}) gives as the adjusted values $q$ = $r$ = 5.192. This indicates a purely nonextensive regime, better described by Eq. (\ref{qE7}), without a statistical crossover. 

\begin{figure}[!htb]
\centering
\includegraphics[width=7.8cm]{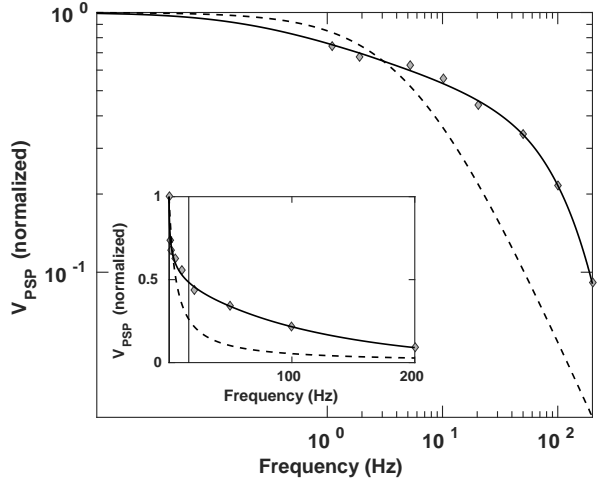}
\caption{Electrophysiological data recorded with extracellular field potential technique. Fitting of the excitatory field $V_{PSP}$ of hippocampal synapse adapted from Fig. 1B of \cite{kilbride}. Full lines represent fitting with Eq. (\ref{qE9}) whereas dashed lines were fitted with Eq. (\ref{qE5}). The inset shows the linear-linear representation of data. The vertical line on the inset marks the crossover frequency at 16.026 Hz. The first crossover at 0.182 Hz is not visible.}
\label{figure2}
\end{figure}

 We also hypothesized whether nonextensivity could be verified in non-mammalian synapses by investigating $V_{PSP}$ data from the avian auditory system (Fig. \ref{figure1}, Bottom) \cite{cook}. Transitions from nonextensivity ($q$ = 4.326) to extensivity ($r$ = 1.000) are observed with first and second crossover frequencies given by $f^{*}_{q}$ = 1.467 Hz and $f^{*}_{r=1}$ = 75.165 Hz, respectively. Next, we investigated data from the hippocampus \cite{kilbride} (Fig. \ref{figure2}), which give as estimated parameters $q$ = 7.933 and $r$ = 1.013 with $f^{*}_{q}$ = 0.182 Hz and $f^{*}_{r=1}$ = 16.026 Hz. Here we used Eq. (\ref{cross1_r1}) and Eq. (\ref{cross2_r1}) to calculate approximate values for the crossover frequencies. 

Our results are summarized in table \ref{table1}. Adjustments to the $1/f$-based equation, Eq. (\ref{qE5}), were only partially achieved in all cases.  We conclude: (a) there are statistical transitions in STD phenomena; (b) nonextensivity is present in auditory system synapses of mammalian and non-mammalian species; (c) nonextensivity is observed in data from intracellular and extracellular environments; (d) nonextensivity in dentate gyrus plasticity suggests a relation between nonextensivity and the neuronal substrate involved in learning and memory of other hippocampal areas; (e) diversified statistical transitions point out that although neurotransmitter secretion has similar machinery, fine structural and functional aspects of each synapse may dictate significantly the type of statistical transition \cite{regher}. 

\begin{table}[h]
\centering
\caption{Summary of simulations with the best fitted parameters obtained using equations Eq. (\ref{qE5}),  Eq. (\ref{qE9}), Eq. (\ref{qE12}) and Eq. (\ref{qE13}). Time constants, $\tau$ were extracted from \cite{henrique2,neher2,trommershauser,cook,kilbride}. Quantal sizes, $Q$ from \cite{neher3,truss,biro}.}
\begin{tabular}{@{}lccccccccc@{}}
\toprule
Reference    & q    &$\lambda$ ($s$) &$\mu$ ($s$) &r  &$\tau$ (s) &$\kappa$ ($s^{-1}$) &$Q$  ($arb. unit$) &$p$ \\
\midrule
Fig. \ref{figure1}, Top    & 5.192 & 3.989  & 0.022  & 5.192  & 4.2  &42.933 &40.0   & 0.024\\
Fig. \ref{figure1}, Bottom & 4.326 & 0.205  & 0.004  & 1.000  & 1.1  &45.682 &36.5   & 0.005\\ 
Fig. \ref{figure2}         & 7.933 & 0.790  & 0.009  & 1.013  & 8.0  &10.847 &40.0   & 0.002\\
\bottomrule    
\end{tabular}
\label{table1}
\end{table}

\section{Discussion}

Applications of nonextensive statistics are not documented in brain synapses, although they have been reported in many other systems. To address this issue, motivated by a limitation of the $1/f$-model to describe STD, we are proposing a new theoretical approach to reveal both nonextensivity and possible statistical transition embedded in this type of plasticity. Influenced by a paradigmatic nonextensive differential equation, we developed a nonlinear model to be applied to electrophysiological recordings from brain synapses. In contrast with the $1/f$-model, our results agree with empirical data, providing a better adjustment in the higher frequency stimulation range. A remarkable advantage of our proposal is the simplicity to test long-range correlations and statistical crossover associated with the STD phenomenon. Moreover, the results give additional support against a synaptic transmission purely ruled by random mechanisms. Although we recognized that our model still represents an oversimplified description of STD mechanisms, it preserves a manner to study long-range correlations neglected in other theoretical descriptions. Therefore, a complete model, allowing a rigorous conjunction of theoretical construction and physiological mechanisms, is still required to permit verification of nonextensivity in a more realistic physiological environment. In this sense, calcium contribution, vesicle replenishment, and receptor desensitization are fundamental requirements to expand this seminal model to a general STD description. 

The obtained $q$-indexes values did not exhibit the confined range of $1 \leq q < 3$ as reported in our study carried out at the neuromuscular junction \cite{adr1}. In this previous work, we adopted a $q$-Gaussian distribution in order to verify the presence of nonextensivity from spontaneous miniature potentials. We used a $q$-Gaussian distribution that admits $q$-index values within the range of $1 \leq q < 3$. However, $q$-exponential and $r$-exponential functions do not require such restriction. For instance, $q > 3$ was evaluated in psychophysical data, image analysis, perceptual computing, and detection and location of mean level-shifts in noise \cite{kimura,rickard,nie,pacheco}. In our case, we attribute the existence of $q > 3$ or $r > 3$ to a relaxation process already suggested in studies involving stock markets and solar winds at the distant heliosphere, respectively \cite{pavlos,pavlos2,burlaga}. In fact, high values for the $q$-index are associated with a relaxation process corresponding to a metastable state. In this sense, we hypothesized that high $q$-index values arise due to statistical transitions and relaxation in the STD mechanism.  During  stimulus application, vesicles placed at AZ are found to be in a balance between a stable and an unstable state. In this sense, Long \textit{et al.} \cite{long} suggested a metastable state regulating the vesicle fusion into a hemifusion process. In this framework, specific proteins may also participate from SNARE complex as indicated by Tang \textit{et al.} \cite{sudhof2}.

The synaptic ending is a propitious biological system to observe the existence of nonextensivity due to its peculiar ultrastructural features \cite{satzler}. For instance, at the calyx of Held, AZ area is 0.1 $\mu m^{2}$, with $2$ docked vesicles per AZ in a terminal volume of 480 $\mu m^{3}$, while at the hippocampal button, AZ area is 0.039  $\mu m^{2}$ with $10$ docked vesicles per AZ in a terminal volume of 0.08 $\mu m^{3}$ \cite{stevens97}. From such morphology, one may presume that a smaller volume and a higher number of docked vesicles in hippocampal synapses constitute physiological substrates consistent with higher $q$-indexes, as compared to giant auditory synapses. Indeed, in restricted spatial dimensions, vesicle fusion on AZs can influence the remaining vesicles to get a probability to be dragged in a multiquantal release or even inhibiting the nearest vesicle to fuse with the terminal \cite{bennett3}. During the early stimulation phase, the readily releasable pool is mobilized. However, further exocytosis, in response to sustained stimulus, leads to depletion of the readily available pool and recruitment of the other pools of vesicles. This non-uniform or heterogeneous neurotransmission is also supported by evidence of physical interactions among vesicles on the same AZ STD \cite{harlow}.A theoretical explanation for the heterogeneous framework for STD was achieved by Trommershauser \textit{et al.} assuming two $p$ classes. They associated a high $p$ to the readily releasable vesicles, released during the early stimulus, and low $p$, for those fusing at higher stimulation levels \cite{trommershauser}.These assumptions led the authors to study previous experiments from Gersdoff \textit{et al.}, whose theoretical modeling are in agreement with empirical results \cite{henrique2}. However, they do not consider physical interactions between vesicles as a source for a heterogeneous STD mechanism. 

In the present work, we suggest nonextensivity and statistical crossover as important factors to explain STD heterogeneity. Heterogenous synapses, represented by statistical transitions, guarantee fidelity over the transmission of a broad range of stimulation without abolishing the postsynaptic response. Since $q > 1$ reflects fractality, our results show that STD presents a fractal behavior not previously described in other reports. A correspondent physiological environment for $r$ = 1 in auditory giant synapses and hippocampus can be interpreted using the morphological argument discussed above. As it is well known, higher frequencies promote a decrement of $p$ by exhaustion of the readily releasable pool, accelerating the recruitment of  vesicles from other storages. If we consider that electrical stimulations promote neural swelling, vesicle traffic facilitation is expected from these storages due to the increment of the intracellular milieu size \cite{tasaki}. High-frequency stimulus can also accelerate the metabolism, decreasing the physical interaction likelihood on each AZ. Combined, both aspects are arguments for a transition from a nonextensive to an extensive behavior. Altogether, we advocate that, despite similarities in exocytosis mechanisms shared by different synapses, structural and functional elements inherent to each terminal reflect STD statistical properties and nonextensivity degree.

\section{Conclusion}

To the best of our knowledge, this is the first work that assesses a nonextensive behavior in brain synapses. In this framework, our main concern was formulating a physiological model to uncover nonextensivity and possible statistical crossover in synaptic transmission. Despite functional and morphological particularities involved in each synapse here studied, they share common features in their exocytosis machinery dynamics such as a similar electrophysiological response. Both remarks encourage us to study how ubiquitous is the nonextensivity in synapses by examining three different brain regions. We found a consistent nonextensive scenario for mammalian and non-mammalian synapses, different species, brain areas and intracellular as well as extracellular compartments. In fact, the results validated the Tsallis theory, at least in auditory and cortical neurons, evidencing that synaptic transmission is not governed only by random mechanisms. Beyond that, statistical transitions arising as a function of stimulus level provide a novel and additional evidence in favor of statistical heterogeneity in neurotransmitter release. Altogether, these findings represent an important step toward the elaboration of more realistic models of STD mechanisms based on nonextensive formalism. They also reinforce the richness and complexity of neuroplasticity phenomena. Lastly, we hope to elaborate experimental protocols for acquisition of our own data for further test in a rigorous model taking into account a more realistic physiological scenario. 

\section*{Acknowledgements}
The authors thank Constantino Tsallis for his valuable suggestions and discussions.

\section*{Compliance with Ethical Standards}
\textbf{Conflict of interest} The authors declare that there is no conflict of interest associated with this publication.


\bibliography{bibmanuscriptSTD210617}
\bibliographystyle{nar}

\end{document}